\documentclass[prb,twocolumn,showpacs,preprintnumbers,amsmath,amssymb]{revtex4}

\usepackage{graphicx}
\usepackage{bbm}
\usepackage{float}

\begin{document}

\preprint{}
\title{The ferroelectric transition in YMnO$_3$ from first principles}
\author{Craig J. Fennie and Karin M. Rabe}
\affiliation{Department of Physics and Astronomy, Rutgers University,
        Piscataway, NJ 08854-8019}
\date{\today}

\begin{abstract}
We have studied the structural phase transition of multiferroic YMnO$_3$ 
from first principles. Using group-theoretical analysis and 
first-principles density functional calculations of the total energy and
phonons, we perform a systematic study of the energy surface around the prototypic 
phase. We find a single instability at the zone-boundary which 
couples strongly to the polarization. This coupling is the mechanism
that allows multiferroicity in this class of materials.
Our results imply that YMnO$_3$ is an improper ferroelectric. 
We suggest further experiments to clarify this point.

\end{abstract}


\maketitle


The hexagonal manganites, $Re$MnO$_3$ for $Re$=Ho-Lu and Y, are a class of multiferroic 
materials that are simultaneously ferroelectric (FE) and antiferromagnetic 
(AFM).~\cite{abrahams.acta.01} In particular, YMnO$_3$ is FE at room temperature (RT) 
crystallizing in space group P6$_3$cm (C$_{6v}$, $Z$=6).~\cite{vanaken.acta.01} 
Above $\approx$1270K it has been shown to undergo a transition to paraelectric (PE) 
P6$_3$/mmc (D$_{6h}$, $Z$=2).~\cite{luka.ferro.74,nenert}

The sequence of phase transitions from the low temperature FE to the high temperature PE
phase has been the subject of much debate. Based on pyroelectric measurements and lattice 
constants determined by x-ray diffraction, Ismailzade and Kizhaev argued that YMnO$_3$ 
undergoes a FE transition from P6$_3$cm ($Z$=6) to an intermediate nonpolar P6$_3$/mcm 
($Z$=6) at $\approx$930K.~\cite{ismailzade.spss.65} In contrast, Lukaszewicz and 
Karut-Kalicinska were unable from x-ray diffraction to resolve this intermediate non-polar
space group~\cite{luka.ferro.74} at 1005 K and saw no clear anomalies in the lattice constants
at $\approx$930K. Katsufuji et al$.$ confirmed the latter results on the temperature 
dependence of the lattice constants and resolved the FE space group up to 1000K, the highest 
temperature they considered.~\cite{katsufuji.prb.02} The FE space group immediately below the 
zone-tripling transition is also supported by the structural refinements of Lonkai et al$.$ 
on the closely related compounds ReMnO$_3$, for Re = Lu, Tm, Yb.~\cite{lonkai.prb.04}
On the other hand, N\'{e}nert et al$.$ recently interpreted their x-ray diffraction measurements 
to support the observation of the intermediate P6$_3$/mcm ($Z$=6) between 1020K and 
1270K.~\cite{nenert} 
The presence of some kind of feature at 930K is suggested by small anomalies in resistivity 
data at $\approx$ 950 K~\cite{katsufuji.prb.01} (also observed in HoMnO$_3$~\cite{sang.unpublished}). 
Thus, the nature of the phase right below the 1270 K transition out of the PE phase, a 
precise characterization of the behavior around 900-1000 K, and the identification of 
the PE-FE transition in this material, remain subjects of great interest.

It has been established that to coexist with magnetism, the nature of the
ferroelectricity cannot be that found in prototypical perovskite FEs
such as BaTiO$_3$.~\cite{hill.jpchemB.00} Indeed, first-principles
calculations have shown that the usual indicators of a FE instability
(e.g. large Born effective charges) are absent in
YMnO$_3$.~\cite{vanaken.natmat.04} Polarization {\bf P}$_s$ = 6 $\mu C/cm^2$
results from the tilting of the Mn-centered
oxygen octahedra and buckling of the Y-O planes, with no significant
off-centering of the Mn cations as would be characteristic of a perovskite
FE.~\cite{vanaken.natmat.04} As such, YMnO$_3$ has been referred to as a ``geometric
ferroelectric.''~\cite{spaldin.natmat.04}
In this paper, we combine a group theoretical
analysis of the experimental FE structure with first-principles calculations
to describe the mechanism by which this ``geometric ferroelectricity'' originates.
Computation of the coupling between a zone-boundary instability and a soft,
but stable, zone-center phonon shows that it induces a low temperature {\bf P}$_s$
$\approx$ 6.5 $\mu C/cm^2$ consistent with experiments. Identifying
this zone-boundary mode as the primary order parameter (OP), we argue that
there is an improper FE transition in YMnO$_3$
at the observed zone-tripling transition, $\approx$ 1270K.
We suggest possible origins for the feature that has been observed around 1000K.

First-principles density-functional calculations using projector augmented-wave 
potentials were performed within the LSDA+U (local spin-density approximation 
plus Hubbard U) method~\cite{anisimov.prb.91} as implemented in the {\it Vienna ab initio 
Simulation Package} ({\sf VASP}).~\cite{VASP,PAW} The wavefunctions were expanded
in plane waves up to a kinetic energy cutoff of 500 eV. Integrals over the Brillouin
zone were approximated by sums on a $4 \times 4 \times 2$ $\Gamma$-centered $k$-point mesh.
Values of on-site Coulomb interaction U = 8eV and exchange parameter J=0.88eV 
were used for the Mn d-orbital as calculated by Ref.~\onlinecite{medvedeva.jpcm.00}.
Polarization was calculated using the modern theory of polarization~\cite{king-smith.prb.93} as
implemented in {\sf VASP}. All results presented are within LSDA+U with $A$-type AFM 
order.~\cite{vanaken.natmat.04} Calculations of phonons, structural energetics, 
and polarization were also performed within LSDA, and with both 
frustrated AFM~\cite{medvedeva.jpcm.00} and FM order; the results are essentially 
unchanged. 

We performed full optimization of the lattice parameters and internal
coordinates in the reported FE space group P6$_3$cm.  As can be seen in 
Table~\ref{table:wyckoff}, excellent agreement with the
experimental RT structure (lattice constants within $\approx$0.5$\%$ and bond 
lengths within $\approx$1$\%$ of the experimental values) is obtained.

\begin{table}[t]
\caption{Crystal structure of ferroelectric YMnO$_3$,
Space Group: $P$6$_3$$cm$, Exp$.$ room temperature: $a$= 6.139$\AA$, \,$c$= 11.41$\AA$.
LSDA+U: $a$= 6.099$\AA$, \,$c$= 11.42$\AA$ }
\begin{ruledtabular}
\begin{tabular}{lclc}

Atom &Exp. Ref.\onlinecite{vanaken.acta.01}& &LSDA+U\\ \hline

\begin{tabular}{l}Y$_1$\,(2a)\\Y$_2$\,(4b)\\Mn\,(6c)\\
O$_{\mathrm{ap}(1)}\,(6c)$\\O$_{\mathrm{ap}(2)}\,(6c)$
\\O$_{\mathrm{eq}(3)}$\,(2a)\\O$_{\mathrm{eq}(4)}$\,(4b)\\ \end{tabular}

&\begin{tabular}{llllr} 0&&0&&0.2743\\$1\over3$&&$2\over3$&&0.2335\\0.3352&&0&&0.0000\\
0.3083&&0&&0.1627\\0.3587&&0&&-0.1628\\0&&0&&-0.0218\\
$1\over3$&&$2\over3$&&0.0186\\ \end{tabular}
                   &
&\begin{tabular}{llllr} 0&&0&&0.2751\\$1\over3$&&$2\over3$&&0.2313\\0.3334&&0&&0.0000\\
0.3058&&0&&0.1642\\0.3601&&0&&-0.1641\\0&&0&&-0.0242\\
$1\over3$&&$2\over3$&&0.0207\\ \end{tabular}

\end{tabular}
\end{ruledtabular}
\label{table:wyckoff}
\end{table}

\begin{figure}[t]
\includegraphics[scale=0.3]{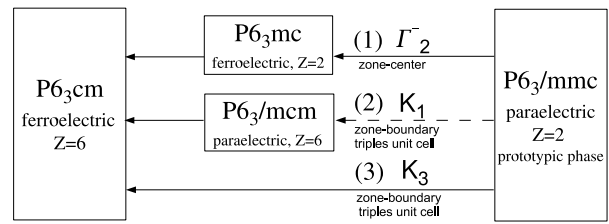}\\
\caption{\label{fig:groups} Group-subgroup sequence of allowed phase transitions
connecting the high-temperature prototypic phase, P6$_3$/mmc, to the low-temperature
ferroelectric phase, P6$_3$cm. 
Each sequence is labeled by an irrep of the prototypic phase where structural 
distortions associated with these irreps determine possible intermediate 
phases.}
\end{figure}

Group theoretical analysis (e.g. Ref.~\onlinecite{isotropy,bilbao}) shows that
there are three distinct symmetry-allowed phase transition sequences connecting 
the high-temperature prototypic phase, P6$_3$/mmc, to the 
low-temperature FE phase, P6$_3$cm, as shown in Fig.~\ref{fig:groups}. There is 
a clear experimental consensus that the zone-tripling transition occurs before 
or at the FE transition, thereby excluding P6$_3$mc as an intermediate phase and 
ruling out path (1). In principle, path (2) could account for both the observed
zone-tripling transition at 1270K and a possible second transition at $\approx$1000K.  
This transition sequence would be a first-order transition to the non-polar intermediate
phase (by the freezing-in of a $K_1$ phonon), followed by a second-order, proper FE 
transition to the RT phase (due to a softening of an infrared-active phonon).
Path (3) describes a second-order transition out of the prototypic phase, by the
freezing-in of a $K_3$ phonon, directly to the FE space group. Unlike what has been
proposed in the literature~\cite{lonkai.prb.04}, Path (3) does not require a second 
transition to account for FE order (as discussed below).~\cite{stokes.pt.91}

Independent of the path,~\cite{perez-mato.prb.04} the distortion that relates the prototypic
phase to the low temperature FE phase can be decomposed into the symmetry-adapted modes of the 
prototypic phase as follows: $\Gamma^-_2$ (zone-center polar mode), $K_1$ and $K_3$ 
(zone-boundary modes at $q$=$1\over3$,$1\over3$,0), and $\Gamma^+_1$ (the identical 
representation).~\cite{perez-mato.jpc.81,manes.prb.82} 
This decomposition is shown in Table~\ref{table:decomp} for the experimental 
RT structure. The excellent agreement between theory and experiment for the structure,
Table~\ref{table:wyckoff}, results in similar agreement for the  mode decomposition.

The relative strength of each mode in the FE structure may be quantified by considering
its amplitude,~\cite{comment.amplitude} $Q$, also given in Table~\ref{table:decomp}. We find, 
as did previous first-principles calculations and structural analysis,~\cite{vanaken.natmat.04}
that the FE structure is dominated by the $K_3$ mode, with Y and O$_{eq}$ atoms undergoing 
the largest displacements. The relative sizes of the symmetry-adapted mode amplitudes in 
the RT FE phase offer a valuable clue to which of the three phase-transition paths is taken 
by YMnO$_3$. In soft-mode driven structural transitions one expects a mode that stabilizes 
an intermediate phase at higher temperatures to be at least comparable in strength (not 
necessarily the largest) to any other modes appearing in the lower temperature phase. 
If $Q_{K_1}$ were the primary structural distortion out of the prototypic phase, it should be 
retained through the second transition into the low temperature FE phase, which is not the 
case in YMnO$_3$. Thus path (2) tends to be ruled out by the smallness of $Q_{K_1}$.
Finally, we consider the remaining possibility, path (3). The fact that $Q_{K_3}$ is
one to two orders of magnitude larger than $Q_{\Gamma^-_2}$ and $Q_{K_1}$ respectively,
suggests that $Q_{K_3}$ is in fact the primary structural distortion out of the prototypic
phase (e.g Ref.~\onlinecite{perez-mato.prb.04}) and thus that YMnO$_3$ follows path (3).

\begin{table}[b]
\caption{Decomposition of Exp$.$ Ref.~\onlinecite{vanaken.acta.01} atomic displacements (\AA) into
symmetry-adapted modes of prototypic $P$6$_3$/$mmc$. Mode amplitudes, $Q$, are shown in \AA. }
\begin{ruledtabular}
\begin{tabular}{lc}

\begin{tabular}{l} 
\\ \\Y$_1$\\Y$_2$\\Mn\\O$_{\mathrm{ap}(1)}$\\O$_{\mathrm{ap}(2)}$
\\O$_{\mathrm{eq}(3)}$\\O$_{\mathrm{eq}(4)}$\\ $Q_{exp}$\\$Q_{theory}$ \end{tabular}

&
\begin{tabular}{rcrcr}
\begin{tabular}{c}$\Gamma^-_2$ \\ $[001]$\\\hline -0.038\\-0.038\\-0.005\\
-0.005\\-0.005\\\,\,0.054\\\,\,0.054\\ 0.16\\0.19\end{tabular}
&&
\begin{tabular}{c} $K_1$ \\$[100]$\\\hline0\\0\\0.011\\0.001\\0.001\\0\\0\\ 0.03\\0.01\end{tabular}
&&
\begin{tabular}{cc} \multicolumn{2}{c}{$K_3$}\\$[100]$& $[001]$ \\ \hline
0&\,\,0.310\\0&-0.155\\0&0\\-0.155&0\\\,\,0.155&0\\0&-0.307\\0&\,\,0.153\\
\multicolumn{2}{c}{0.93}\\ \multicolumn{2}{c}{1.00} \end{tabular}
\end{tabular}

\end{tabular}
\end{ruledtabular}
\label{table:decomp}
\end{table}

First principles ($T$=0) investigation of the phonons and the structural
energetics around the prototypic phase also shows path (3) to be the most
likely scenario. For the structural parameters of the prototypic phase, we 
used the minimum energy structure within P6$_3$/mmc: $a$=6.168$\AA$, 
$c$=11.23$\AA$ and $u_{O_{ap}}$=0.9166.  This corresponds to a nearly homogeneous 
compression of the lattice constants by about 1.4$\%$ compared with those measured by 
Lukaswewicz at 1280K, an underestimate typical of LSDA calculations. Phonons at the 
$\Gamma$ and $K$-point in the Brillouin zone of the prototypic phase were computed by 
constructing the relevant block of the dynamical matrix from Hellmann-Feynman forces. 
This allowed us to greatly reduce the number of calculations and provided higher numerical 
accuracy in the determination of the force-constant eigenvectors. At the $\Gamma$-point
we find that the three infrared-active $\Gamma^-_2$ phonons are stable with the lowest
frequency $\Gamma^-_2$(TO1): $\omega$ = 90 cm$^{-1}$. At the  $K$-point, we computed the 
three $K_3$ phonons by a similar method and found one highly unstable mode $K_3$(1): $\omega$ 
= i 153 cm$^{-1}$. In contrast, both $K_1$ phonons are quite stable with $K_1$(1): 
$\omega$ = 264 cm$^{-1}$ and $K_1$(2): $\omega$ = 522 cm$^{-1}$.  Since lattice 
instabilities are known in some cases to be sensitive to volume, we repeated these 
calculations at the experimental lattice constants; the results are qualitatively unchanged.

The stability of the $\Gamma^-_2$ and $K_1$ modes is additional evidence to rule out 
paths (1) and (2), respectively. Admittedly, as these are $T$=0 
calculations there is still a possibility that an intermediate phase could be stabilized by an anomalous
entropic contribution to the free energy at finite temperatures,~\cite{dove} so that, e.g$.$ 
P6$_3$/mcm would be more stable than P6$_3$cm for a range of temperatures below 1270K. 
However, in light of the strength of the $K_3$ instability and as we will show, the associated energy scale, 
the transition out of the PE phase is most naturally driven by the $K_3$ mode, corresponding to path (3).

\begin{figure}[t]
\includegraphics[scale=0.18]{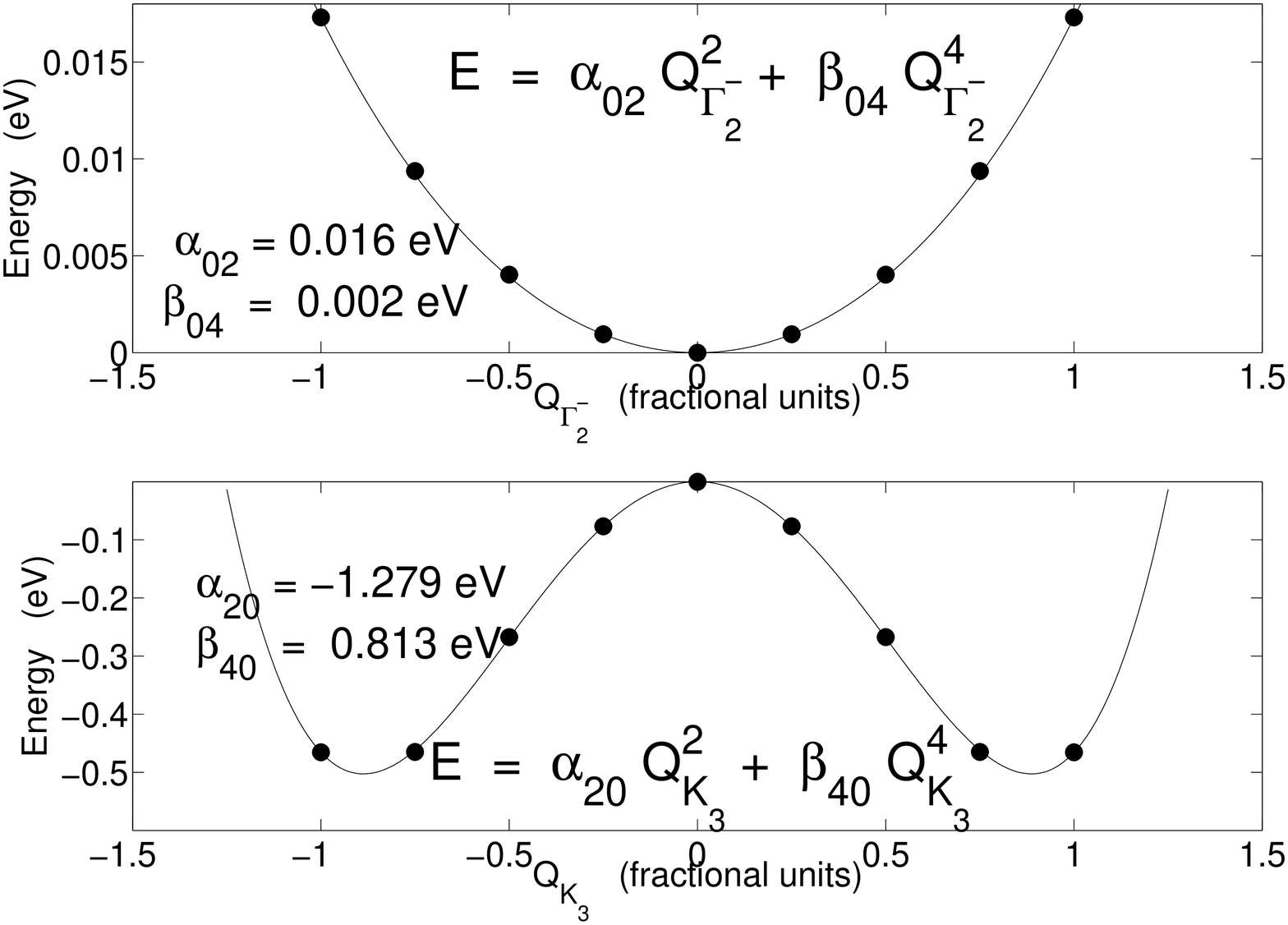}\\
\caption{\label{fig:energy_K_G} Energy as a function of (top) $Q_{\Gamma^-_2}$
and (bottom) $Q_{K_3}$. }
\end{figure}

We can understand how the $K_3$(1) zone-boundary instability produces 
a polarization by calculating the energy surface around the prototypic 
phase as a function of the relevant modes. To simplify the analysis, the 
fact that all $\Gamma^+_1$ and $K_1$ phonons are stable in the prototypic phase 
and contribute very little to the total structural distortion suggests that 
the coupling of these modes to $K_3$(1) will be small and can be ignored. 
Within this reduced subspace we expand the energy of the crystal including 
all symmetry allowed terms to fourth order in $Q_{K_3}$ and $Q_{\Gamma^-_2}$ 
to obtain:
\begin{eqnarray} \mathcal{F}(Q_{K_3},Q_{\Gamma^-_2}) &=& \alpha_{20} Q_{K_3}^2 +
\alpha_{02} Q_{\Gamma^-_2}^2 +\beta_{40} Q_{K_3}^4 +\beta_{04} Q_{\Gamma^-_2}^4 
\nonumber \\
& & +\beta_{31} Q_{K_3}^3  Q_{\Gamma^-_2} + \beta_{22} Q_{K_3}^2  Q_{\Gamma^-_2}^2
\label{eq:free} \end{eqnarray}

\begin{figure}[t]
\includegraphics[scale=0.18]{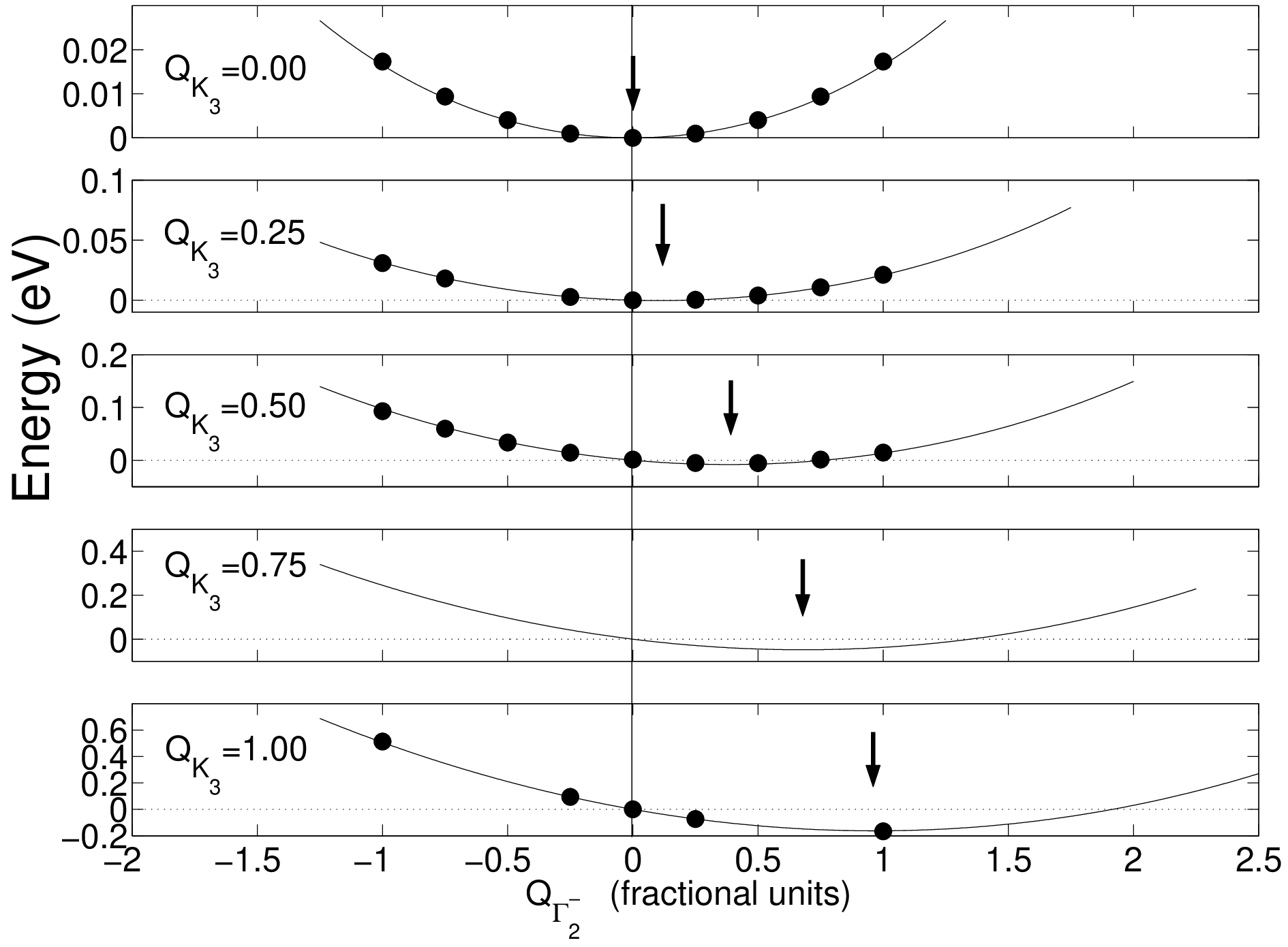}\\
\caption{\label{fig:energy_fixed_K} Energy as a function of $Q_{\Gamma^-_2}$
at fixed $Q_{K_3}$.}
\end{figure}

Next we performed a series of first-principles calculations where we 
freeze-in various amplitudes of $Q_{K_3}$ and $Q_{\Gamma^-_2}$ and 
compute the total energy. The energy expansion, Eq.~\ref{eq:free}, was then 
fit using only six data points, yielding excellent agreement with the 
remaining $\sim$20 data points. As we will show, the model energy fits the 
calculated data well over the entire energy surface, justifying our 
truncation of the Taylor expansion to fourth order. In Fig.~\ref{fig:energy_K_G} 
we show the total energy (of the 30-atom cell) as a function of the $Q_{\Gamma^-_2}$ 
(top) and $Q_{K_3}$ (bottom) fractional displacements. As expected, 
the $\Gamma^-_2$ mode is stable while the double-well potential 
of the $K_3$ mode is apparent. The coupling between the modes can  
be seen by plotting the energy as a function of $Q_{\Gamma^-_2}$ at fixed 
$Q_{K_3}$ as in Fig.~\ref{fig:energy_fixed_K}.
Notice that once $Q_{K_3}$ becomes non-zero the equilibrium position of 
$Q_{\Gamma^-_2}$ (indicated by the arrow) shifts to a positive value and continues
to increase with increasing $Q_{K_3}$. Also as $Q_{K_3}$ increases, the 
curvature of the energy increases due to the $\beta_{22}$=0.155 eV term 
which renormalizes the quadratic coefficient of $Q_{\Gamma^-_2}$. The effect 
of the $K_3$ mode is not to decrease the stability of the $\Gamma^-_2$ mode, 
but rather to ``push'' the $\Gamma^-_2$ mode to a non-zero equilibrium 
position, playing a role analogous to that of a field.
This ``geometric field'' is the mechanism for ferroelectricity in the 
hexagonal manganites.~\cite{comment.sic} Further, the strong $\beta_{31}$=-0.334 eV coupling 
between the $K_3$(1) and $\Gamma^-_2$(TO1) {\it lattice} modes, i.e. phonons,
is responsible for the comparatively large value of {\bf P}$_s$ $\approx$ 
6$\mu C/cm^2$. Indeed, if this coupling were to be turned off, the 
polarization generated by $K_3$(1) (coupling only to the electronic density) 
is reduced to {\bf P}$_s$ $\approx$ 0.1$\mu C/cm^2$.~\cite{fennie.unpublished}

Now let us return to the discussion of the PE-to-FE path. If we now
identify $K_3$(1) as the primary OP and $\Gamma^-_2$(TO1) as a secondary
OP, inspection of Fig.~\ref{fig:energy_fixed_K} reveals interesting crossover
behavior of the $\Gamma^-_2$ mode.  Minimizing the free energy over 
$Q_{\Gamma^-_2}$ leads to two cases:
                                                                                                
Region 1: $Q_{K_3} << \sqrt{|{\alpha_{02} \over \beta_{22}} |}$\,\,
$\Longrightarrow$\,\, $Q_{\Gamma^-_2} \propto -{\beta_{31} \over 2\alpha_{02}} Q_{K_3}^3
$
                                                                                                
Region 2:  $Q_{K_3} >> \sqrt{|{\alpha_{02} \over \beta_{22}} |}$\,\,
$\Longrightarrow$ \,\,$ Q_{\Gamma^-_2} \propto -{\beta_{31} \over 2\beta_{22}}  Q_{K_3}
$

\begin{figure}[t]
\includegraphics[scale=0.18]{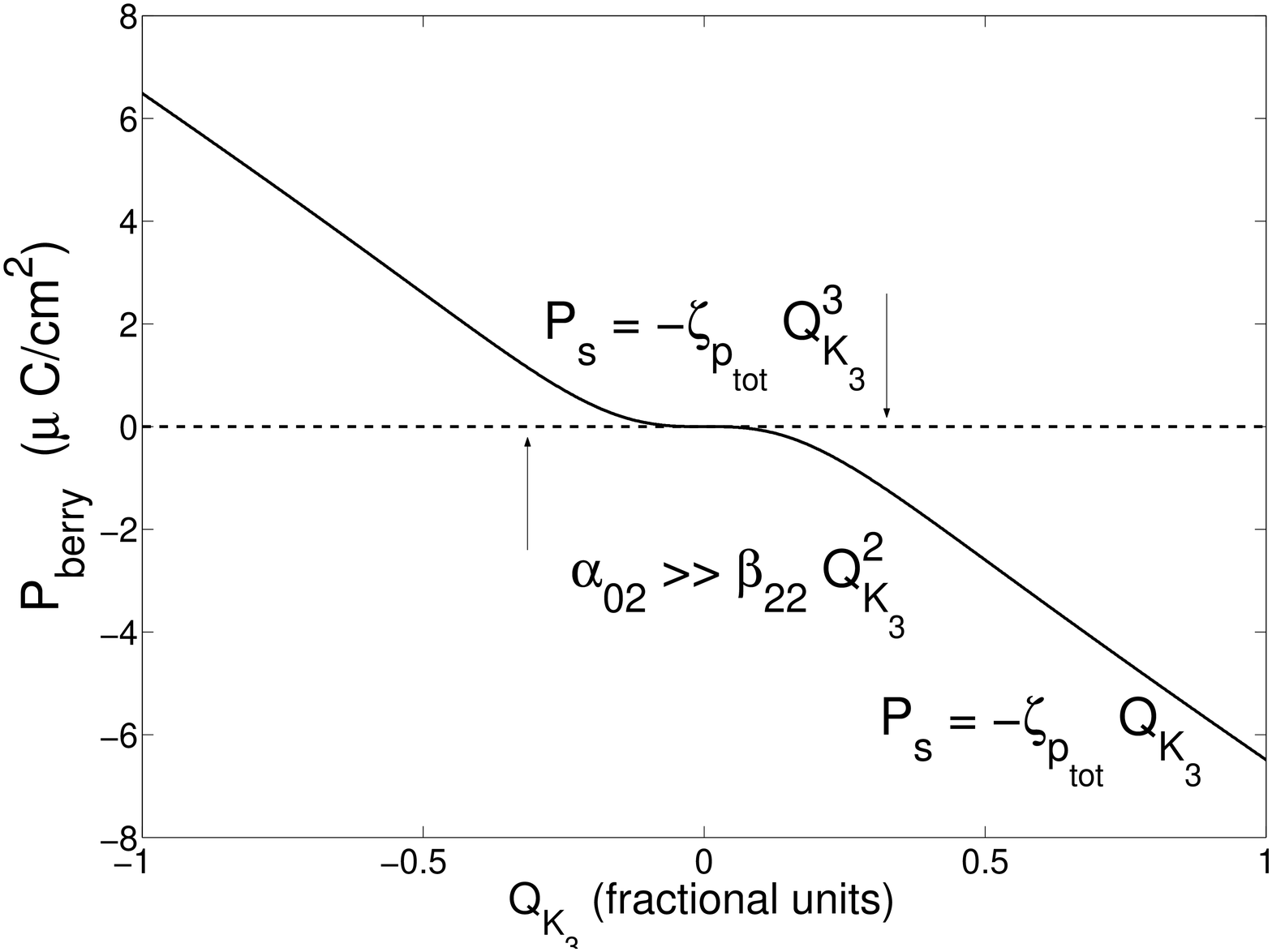}\\
\caption{\label{fig:polarization} Polarization, {\bf P}$_s$, vs. primary order parameter,
$Q_{K_3}$.}

\end{figure}

Since the polarization, {\bf P}, is a first order polar tensor it has the 
same transformation properties as the symmetry-adapted $\Gamma^-_2$ mode, i.e.
${\bf P} \propto \, Q_{\Gamma^-_2}$. Using the modern theory of 
polarization we can calculate the polarization as a function of the 
structural $\Gamma^-_2$ mode. The physical quantity of interest though is the 
polarization as a function of the primary OP, $Q_{K_3}$, which we show in 
Fig.~\ref{fig:polarization}.  As the $K_3$ structural distortion is initiated,
{\bf P}$_s$ becomes non-zero but is small due to the $Q_{K_3}^3$ dependence
until the emergence of a crossover to {\bf P}$_s$ $\propto$ $Q_{K_3}$. 
This leads to an apparent ``turn-on'' at a finite value
of $Q_{K_3}$. We calculate the total energy (per formula unit) 
lowering from the PE to FE to be $\mathcal{O}$(1000K) (the precise value 
fortuitously being 1240K) and the width of this crossover region to be $\mathcal{O}$(100K).

In this paper, we have shown that the overwhelming weight of the experimental and theoretical 
evidence supports path (3), with an improper FE transition out of the high-temperature 
PE phase at 1270 K. The $K_3$(1) phonon is strongly unstable and couples to the $\Gamma^-_2$(TO1)
phonon leading to the observed polarization. The unusual dependence of the polarization on the 
primary order parameter would lead to a broadened pyroelectric peak lower than $T_c$ and could, 
through coupling to strain, lead to an isostructural, FE-to-FE, transition, although this last 
point is only speculative. This differs from previous models~\cite{lonkai.prb.04} in that 
YMnO$_3$ is already polar when this isostructural transition occurs. 
An additional intriguing possibility to explain the apparent lower temperature transition
is that, due to the presumably inhomogeneous 
nature of $K_3$ near $T_c$ (which we have been assuming to be homogeneous),
fluctuations suppress the coupling between $K_3$(1) and $\Gamma^-_2$(TO1) till some 
temperature lower than $T_c$, this second transition temperature likely being
sample dependent.~\cite{hamann} 
In this scenario, the path to the FE phase is still along (3), but the P6$_3$cm phase in the 
intermediate temperature range would be paraelectric in an average sense, as originally suggested
by Lukaszewicz and Karut-Kalicinska.~\cite{luka.ferro.74} 
Clearly, further theoretical and experimental
work is necessary to characterize the lower temperature isostructural transition and 
the nature of the polarization in the intermediate phase.
Finally, it should be appreciated, that regardless of the actual path taken,
the softening of the $K_3$ mode sets up a ``geometric field'' that induces
a substantial polarization, thus providing a mechanism to achieve 
multiferroicity in the hexagonal manganites.

Useful discussions with S-W$.$ Cheong,  D.R$.$ Hamann, G$.$ N\'{e}nert, T$.$ Palstra,
A$.$ Sushkov, and  D.H$.$ Vanderbilt 
are acknowledged. This work was supported by NSF-NIRT Grant No. DMR-0103354.



\begin{thebibliography}{10}
\bibitem{comment.sic}
Using DFT-SIC method, the  first-principles study of 
Ref.~\onlinecite{vanaken.natmat.04} found the total $\Gamma$
structural distortion, i.e. $\Gamma^-_2$ + $\Gamma^+_1$   
independently unstable. It should be emphasized, the essential 
physics of the geometrical field is the coupling between the 
$\Gamma^-_2$ and $K_3$ modes and does not depend on the stability
of the $\Gamma^-_2$ mode.
\end{thebibliography}
\end{document}